\documentclass[10pt,superscriptaddress,twocolumn,amsmath,amssymb,aps,prl,showpacs]{revtex4-1}
\usepackage{mathrsfs}
\usepackage{graphicx}
\usepackage{dcolumn}
\usepackage{bm}
\usepackage{amssymb}
\usepackage{amsmath}
\usepackage{paralist}
\usepackage{float}
\usepackage{mdframed}
\usepackage{booktabs}
\usepackage{mathtools} 
\usepackage{graphicx}
\usepackage[colorlinks,linkcolor=blue,anchorcolor=blue,citecolor=blue]{hyperref} 
\usepackage{cleveref}
\usepackage{url}
\usepackage{color}

\usepackage[all]{xy}

\newcommand{\tmop}[1]{\ensuremath{\operatorname{#1}}}

\def\be{\begin{equation}}
\def\ee{\end{equation}}
\def\bea{\begin{eqnarray}}
\def\eea{\end{eqnarray}}

\begin{document}

\title{Yang-Gaudin model: A paradigm of many-body physics}

\author{Xi-Wen Guan}
\email[]{xiwen.guan@anu.edu.au}
\affiliation{State Key Laboratory of Magnetic Resonance and Atomic and Molecular Physics,
	Wuhan Institute of Physics and Mathematics, APM,  Chinese Academy of Sciences, Wuhan 430071, China}
\affiliation{Department of Theoretical Physics, Research School of Physics and Engineering,
	Australian National University, Canberra ACT 0200, Australia}

\author{Hai-Qing  Lin}
\email[]{haiqing0@csrc.ac.cn}
\affiliation{Beijing Computational Science Research Center, Beijing 100193, China
}
\affiliation{Department of Physics, Beijing Normal University, Beijing, 100875, China}

\date{\today}

\begin{abstract}
Using Bethe's hypothesis, C N Yang exactly solved  the one-dimensional (1D) delta-function interacting spin-1/2  Fermi gas with an arbitrary spin-imbalance   in 1967.
At that time,  using  a different  method, M Gaudin solved the problem of interacting fermions  in a spin-balanced  case.
Later, the 1D delta-function interacting fermion problem was named as the Yang-Gaudin model.
 It has been in general agreed  that a key discovery of C N Yang's work was the cubic matrix equation for the solvability conditions. 
This equation was later independently found by R J Baxter for commuting transfer matrices of 2D  exactly solvable vertex models. 
The equation has since been referred to Yang-Baxter equation, being  the master equation to integrability.
The Yang-Baxter equation  has   been used to solve a wide range  of  1D many-body problems in physics, such as 1D Hubbard model, $SU(N)$ Fermi gases, Kondo impurity problem and  strongly correlated electronic systems etc.
In this paper,  we will briefly discuss  recent developments of the Yang-Gaudin model on several breakthroughs of many-body phenomena,  ranging from the universal thermodynamics to the Luttigner liquid, the spin charge separation, the Fulde-Ferrell-Larkin-Ovchinnikov (FFLO)-like  pairing state  and the  quantum criticality. 
 These developments demonstrate  that  the Yang-Gaudin model has laid out a profound legacy of the Yang-Baxter equation.

\end{abstract}

\pacs{03.75.Ss, 03.75.Hh, 02.30.IK, 05.30.Fk}

\keywords{}

\maketitle

The  Bethe ansatz (BA), i.e. a particular form of wavefunction, was first introduced in 1931 by Hans
Bethe \cite{Bethe:1931} as a way to  obtain the eigenspectrum of the one-dimensional (1D) spin-1/2 
Heisenberg chain.  
 In  Bethe's method, $N!$ plane waves are $N$-fold products of individual exponential phase
factors $e^{\mathrm{i}k_ix_j}$, where the $N$ distinct wave numbers,
$\left\{k_i\right\}$, are permuted among the $N$ distinct coordinates, $x_j$.  Each of
the $N!$ plane waves has an amplitude coefficient in each of
regions, i.e. superposition of all possible plane waves of $N$ particles. 
However, only more than  few decades later, physicists,  L  Hulth\'{e}n \cite{Hulthen}, R. Orbach  \cite{Orbach}, L R Walker    \cite{Walker} , R B Griffith  \cite{Griffith},  J des Cloizeaux and Pearson \cite{Cloizeau}, and few others,  studied the Bethe's method and the  Heisenberg chain in terms of Bethe's solution. 
In the mid-60's,  C N Yang and C P Yang \cite{YY-1,YY-2,YY-3}  presented a systematic study of the BA equations for the Heisenberg spin chain throughout the full range of anisotropic parameter $\Delta$ with  a  presence of magnetic field. 
While they coined Bethe's  method as Bethe's hypothesis. 

In  60's, Bethe's hypothesis  had been proved to be
invaluable for the field of exactly solvable models in statistical
mechanics.
 In 1963, using Bethe's hypothesis,  E Lieb and W Liniger   \cite{Lieb-Liniger} solved the 1D Bose gas with a delta-function interaction. 
In 1967, C N Yang \cite{Yang} solved the 1D delta-interacting Fermi gas with a discovery of the necessary condition for the Bethe ansatz solvability, which is now known as the Yang-Baxter equation, i.e. the factorization condition--the scattering matrix of a quantum many-body system can be factorized into a product of many two-body scattering matrices. 
In the same year, M Gaudin also rigorously derived the BA equations for the spin-$1/2$ Fermi gas with a spin balance \cite{Gaudin}. 
In 1972,  R J   Baxter \cite{Baxter}  independently showed that such a factorization relation also occurred   as the conditions for commuting transfer matrices in 2D  vertex  models in statistical mechanics. 
The Bethe ansatz approach has since  found success in the realm of condensed matter physics, such as the 1D Hubbard model \cite{Lieb-Wu}, SU(N) interacting Fermi gases \cite{Sutherland:1968}, Kondo impurity problems \cite{Andrei:1983}, BCS pairing models \cite{Dukelsky:2004}, strongly correlated electron systems \cite{1D-Hubbard,Korepin,Sutherland-book,Takahashi-b,Wang-book} and spin chain and ladder compounds \cite{Batchelor:2007}, quantum gases of cold atoms \cite{Cazalilla:2011,yangyou,Guan:2013,Batchelor:2016,Mistakidis:2022}, and also many other problems in physics and mathematics. 

The next  significant progress was made  by C N Yang and C  P Yang \cite{Yang-Yang} in 1969  on the finite temperature problem for the Lieb-Liniger Bose gas. 
They showed that the thermodynamics of the model can be determined from the minimisation conditions of the Gibbs free energy subject to the Bethe ansatz equations. 
Later Takahashi showed  \cite{Takahashi:1971,Takahashi:1972} that the Yang-Yang method was  an elegant way to analytically obtain thermodynamics of integrable models, e.g.  1D Heisenberg spin chains, Hubbard model, etc. 
Recent developments on the exactly solvable models in ultracold atoms have  shown \cite{Guan:2013,Zhao,Guan2007,Guan:2013PRL,Yang:2017,He:2020,PhysRevB.101.035149} that Yang-Yang method provides an elegant way to study universal thermodynamics,  Luttinger liquid, spin charge separation, transport properties  and critical phenomena  for a wide range of low-dimensional quantum many-body systems.
 In this short review,  we shall discuss how the exact solution of the Yang-Gaudin model provides a rigorous understanding of such fundamental many-body physics in terms of the legacy  of Yang-Baxter equation and Yang-Yang thermodynamics.  

In this  paper,  we will  present Bethe ansatz solution of the Yang-Guadin model in Section I and will discuss the ground state, spin charge separation, universal thermodynamics and quantum criticality  for the model with a repulsive interaction in Section II.  In section III,  we will briefly review novel quantum phases of pairing and depairing, the Fulde-Ferrell-Larkin-Ovchinnikov (FFLO) pairing correlation, multicomponent Luttinger  liquids and dimensionless ratios  for the Yang-Gaudin model with an attractive interaction.   A brief conclusion and  a short discussion on the  future study of the Yang-Gaudin model  will be presented  in Section IV.


\section{I. The Yang-Gaudin Model}
\label{Section-I} 

The Hamiltonian for the 1D  contact interacting fermion  problem  \cite{Yang,Gaudin} 
\begin{equation}
H=-\frac{\hbar^{2}}{2m}\sum_{i=1}^{N}\frac{\partial^{2}}{\partial
x_{i}^{2}}+g_{1D}\sum_{1\leq i<j\leq
N}\delta(x_{i}-x_{j}) \label{Ham}
\end{equation}
 describes $N$  fermions of the same mass $m$ with two internal spin  states  confined to a 1D
system of length $L$ interacting via a $\delta$-function potential.
In this Hamiltonian, we denote the numbers of fermions in  the two  hyperfine levels $|\uparrow \rangle $ and $|\downarrow \rangle$  as  $N_{\uparrow}$ and $N_{\downarrow}$, respectively. 
While we denoted the total number of fermions and the magnetization as  $N=N_{\uparrow}+ N_{\downarrow}$ and $M^z=(N_{\uparrow}-N_{\downarrow})/2$. 
The
coupling constant $g_{1D}$ can be expressed in terms of the
interaction strength $g_{1D}=\hbar^{2}c/m$ with  $c=-2/a_{1D}$, where
$a_{1D}$ is the effective 1D scattering length  \cite{Olshanii_PRL_1998}. 
In the following discussion, we let  $2m=\hbar =1$ for our convenience. 
We define  a dimensionless interaction strength $\gamma = c / n$	for our later  physical analysis.
Here the linear density  is defined  by $n = N / L$.  For  a repulsive interaction, $c>0$ and for an attractive interaction, $c<0$. 

Using  Bethe's hypothesis, C N Yang solved the model (\ref{Ham}) with the following   many-body wave function 
\begin{equation}
\psi=\sum_{P}A_{\sigma_{1}\ldots\sigma_{N}}(P|Q)\exp
\textrm{i}(k_{P1}x_{Q1}+\ldots+k_{PN}x_{QN})
\label{wavefunction} 
\end{equation}
for  the domain
$0<x_{Q1}<x_{Q2}<\ldots<x_{QN}<L$. 
Where $\{k_{i}\}$  denote  a set of unequal wave  numbers and  $\sigma_{i}$ with $i=1,\ldots, N$   indicate the spin coordinates. Both $P$ and $Q$ 
are permutations of  indices $\{1,2,\ldots,N\}$, i.e.  $P=\left\{ P_{1},\ldots,P_{N}\right\}$ and $Q=\left\{ Q_{1},\ldots,Q_{N}\right\}$.  The sum runs over all $N!$
permutations $P$ and the coefficients of the exponentials are column
vectors with each of the $N!$ components representing a permutation
$Q$. 
To determine the wave function associated with the irreducible representations of the permutation group $S_N$ and the irreducible representation of the Young tableau for different  up- and down-spin fermions, one first needs to consider the boundary conditions of the wave function, i.e. the continuity of the wave function and discontiuity of  its derivative.  
It follows that the two adjacent  coefficients must meet the two-body scattering relation 
\begin{equation}
A_{\sigma_{1}\ldots\sigma_{N}}(P_{i}P_{j}|Q_{i}Q_{j})=Y_{ij}(k_{Pj}-k_{Pi})
A_{\sigma'_{1}\ldots\sigma'_{N}}(P_{j}P_{i}|Q_{i}Q_{j}),
\end{equation}
where  the two-body scattering matrix $Y_{ij}$  is given by 
\begin{equation}
Y_{ij}(k_{Pj}-k_{Pi})=
\left[\frac{\textrm{i}(k_{Pj}-k_{Pi})T_{ij}+cI}
{\textrm{i}(k_{Pj}-k_{Pi})-c}\right]^{\sigma'_{1}\ldots\sigma'_{N}}_{
\sigma_{1}\ldots\sigma_{N}}
\end{equation}
with  the operator $T_{ij}=-P_{ij}$, here 
$P_{ij}$ is the permutation operator. 

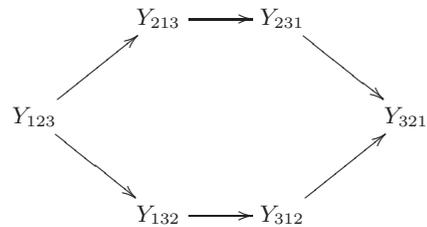
\begin{figure}
\begin{displaymath}
\xymatrix{ & Y_{213} \ar[r] & Y_{231} \ar[dr] & \\
Y_{123} \ar[ur] \ar[dr] & & & Y_{321} \\
 & Y_{132} \ar[r] & Y_{312} \ar[ur] & }
\end{displaymath}
\caption{ The graphical representation of the Yang-Baxter equation Eq. (\ref{YBE}). }
\label{YBE-G}
\end{figure}

The key discovery of C N Yang's work is  that 
the two-body scattering matrix acting on three  linear tensor spaces $V_1\otimes V_2\otimes V_3$
\begin{equation}
Y_{ij}(u)=\frac{\textrm{i}uT_{ij}+cI}{\textrm{i}u-c}
\end{equation}
satisfies the following cubic equation 
\begin{eqnarray}
&&Y_{12}(k_{2}-k_{1})Y_{23}(k_{3}-k_{1})Y_{12}(k_{3}-k_{2})\nonumber \\
&& =Y_{23}(k_{3}-k_{2})Y_{12}(k_{3}-k_{1})Y_{23}(k_{2}-k_{1}),\label{YBE} 
\end{eqnarray}
which has been known as  the Yang-Baxter equation. 
This equation guarantees  no diffraction in the three-particle scattering process, i.e.  $(k'_{1},k'_{2},k'_{3})=(k_{1},k_{2},k_{3})$. 
The Yang-Baxter equation in a graphical representation presents a kind of topological invariance for 
interchanging  the three operators according to the two paths, see Fig.~\ref{YBE-G}.
This equation was immediately seen as a necessary condition to the quantum integrability.

In general, the scattering matrix $Y_{ab}(u)$ satisfies the following relations
\begin{eqnarray}
Y_{ab}(u)Y_{cd}(v) &=& Y_{cd}(v)Y_{ab}(u), \nonumber \\
Y_{ab}(u)Y_{bc}(u+v)Y_{ab}(v) &=& Y_{bc}(v)Y_{ab}(u+v)Y_{bc}(u),\nonumber \\
\nonumber Y_{ab}(u)Y_{ba}(-u) &=& 1.\nonumber 
\end{eqnarray}
Let  us further  introduce the operator $R_{ij}=T_{ij}Y_{ij}$ which acts
on the space $\prod_{n=1}^{N}\otimes V_n$.
Then we define the following monodromy matrix:
\begin{equation}
\mathcal{T}_{N}(u)=L_{N}(k_{N}-u)\ldots
L_{2}(k_{2}-u)L_{1}(k_{1}-u)
\end{equation}
where $L_{i}(k_{i}-u)\equiv R_{i,a}(k_{i}-u)$. 
By  taking  a trance of the
transfer matrix over the auxiliary space $a$, we have 
$\tau(u)=\textrm{Tr}_{a}(\mathcal{T}_{N}(u))$, then we have 
\begin{equation}
\tau(u)|_{u=k_{i}}=\mathfrak{R}_{i}(k_{i}),
\end{equation}
where
\begin{eqnarray}
\mathfrak{R}_{i}(k_{i})&=&R_{i+1,i}(k_{i+1}-k_{i})\ldots
R_{N,i}(k_{N}-k_{i})R_{1,i}(k_{1}-k_{i})\ldots \nonumber\\
&&\times R_{i-1,i}(k_{i-1}-k_{i}).\label{Transfer-matrix}
\end{eqnarray}
The above notations were borrowed  from the quantum inverse scattering method which had  been developed in the 1980's by Faddeev and others
\cite{Korepin}. 

Solving the eigenvalue 
problem of $N$ interacting  particles  (\ref{Ham}) in a periodic box of length $L$, one needs to 
apply the periodic boundary condition
\begin{equation}
\psi(x_{1},\ldots,x_{i},\ldots,x_{N})=\psi(x_{1},\ldots,x_{i}+L,\ldots,x_{N}).
\end{equation}
This is equivalent to solving  the following eigenvalue problem 
\begin{equation}
\mathfrak{R}_{i}(k_{i})A_{E}(P|Q)=\exp(\textrm{i}k_{i}L)A_{E}(P|Q)
\end{equation}
with the transfer matrix $\mathfrak{R}_{i}(k_{i})$ given by (\ref{Transfer-matrix}). 
By some algebraic manipulations, the eigenvalue of the transfer matrix $\tau(u)=\textrm{Tr}_{a}(\mathcal{T}_{N}(u))$ can be straightforwardly obtained. 
Then one can obtain the following C N Yang's  Bethe ansatz equations (BAE) \cite{Yang} for the  Fermi gas 
\begin{eqnarray}
&&{\rm e}^{i k_j L} =\prod_{\alpha =1}^{M} \frac{k_j -\lambda_{\alpha} +ic/2}{k_j -\lambda_{\alpha} -ic/2}, \label{BA1} \\
&&\prod_{j=1}^N  \frac{\lambda_{\alpha}-k_j +ic/2}{\lambda_{\alpha}-k_j -ic/2}=-\prod_{\beta=1}^{M}\frac{\lambda_{\alpha}-\lambda_{\beta}+ic}{\lambda_{\alpha}-\lambda_{\beta}-ic},\label{BA2}
\end{eqnarray}
with $j=1,2,\cdots,N$ and $\alpha=1,2,\cdots,M$. Here $M$ is the number of atoms with down-spins. 
The energy eigenspectrum is given in terms of the quasimomenta $\left\{k_i\right\}$  of the fermions via
$E=\frac{\hbar ^2}{2m}\sum_{j=1}^Nk_j^2$. 
All quasimomenta $\left\{k_i\right\}$  are distinct and uniquely determine the wave function of model Eq.(\ref{wavefunction}).

The fundamental physics  of the model (\ref{Ham}) can be  in principle obtained by solving the transcendental Bethe ansatz equations (\ref{BA1}) and (\ref{BA2}). However,  the  difficulty with this interacting fermion problem lies in finding all the solutions to these Bethe ansatz equations. In the following discussion, we will briefly review recent breakthroughs in the study of the Bethe ansatz solutions of the Yang-Gaudin  model (\ref{Ham}). 

\section{II. Spin Charge Separation and Spin Incoherent Liquid for the Repulsive Fermi Gas}

Finding the solution of the Bethe ansatz equations (\ref{BA1}) and (\ref{BA2}) is cumbersome. In 
the thermodynamic limit, i.e., $L,N \to \infty$, $N/L$ is finite, 
 the above Bethe ansatz equations   can be written as the generalized Fredholm equations 
\begin{eqnarray}
{\rho}(k)&=&\frac{1}{2\pi}+ \int_{-B_2}^{B_2}a_1(k-\lambda){\sigma_1 }(\lambda )d\lambda, \label{BE2-r1}\\
{\sigma_1}(\lambda )&=&\int_{-B_1}^{B_1}a_1(\lambda -k){\rho }(k)dk\nonumber \\
&& - \int_{-B_2}^{B_2}a_2(\lambda-\lambda'){\sigma_1}(\lambda') d\lambda'.\label{BE2-r2}
\end{eqnarray}
The associated   integration boundaries $B_1$, $B_2$   are determined by the relations 
\begin{eqnarray}
n:&\equiv &N/L=\int_{-B_1}^{B_1}{\rho}(k)dk,  \nonumber\\
 n_{\downarrow}:&\equiv& N_{\downarrow}/L=\int_{-B_2}^{B_2}{\sigma_1}(k)dk,\label{repulsive-d}
\end{eqnarray} 
where  $n$ denotes  the linear density while $n_{\downarrow}$ is the density of spin-down Fermions. 
In the above equations we introduced the quasimomentum distribution function $\rho(k)$ and distribution function of the spin rapidity $\sigma_1(\lambda)$ for the ground state.  The boundary $B_1$ characterizes the Fermi point in the quasimomentum space whereas the boundary $B_2$ characterizes the spin rapidity distribution interval with respect to the  polarization.  They can be  obtained  by solving the equations given  in  (\ref{repulsive-d}). 
In the above equations,  we denoted the kernel  function by
\begin{equation}
a_{\ell }(x)=\frac{1}{2\pi}\frac{\ell c}{(\ell c/2)^2+x^2}. \label{a-r}
\end{equation}
The ground state energy per unit length is given by 
\begin{equation}
E=\int_{-B_1}^{B_1}k^2{\rho}(k) d k. \label{E2-r} 
\end{equation}
The ground state energy for weak and strong interactions can be calculated directly from the integral forms of the Bethe ansatz equations (\ref{BE2-r1}) and  (\ref{BE2-r2}), namely 
\begin{eqnarray}
E&=&\frac{1}{12}n^3\pi^2+\frac{1}{2}n^2c+O(c^2),\,\,{\rm for}\, c\ll 1, \label{E-r-wb}\\
E&=&\frac{n^3\pi^2}{3}\left[1- \frac{4\ln2}{\gamma} +\frac{12(\ln2)^2}{\gamma^2} -\frac{32(\ln2)^3}{\gamma^3}\right.\nonumber\\
&&\left. +\frac{8\pi^2\zeta(3)}{5\gamma^3} \right]+O(c^{-4})\,\,{\rm for } \, c\gg 1. \label{E-A-b}
\end{eqnarray}
This  is  a  good approximation  for the spin-balanced Fermi gas with  weakly and  strongly repulsive  interactions. More detailed study of the solutions of the BA equations  was  presented in \cite{Guan:PRA12}. 

In general, for a repulsive interaction, the Bethe ansatz equations (\ref{BA1}) and (\ref{BA2}) only  admit  real roots in the charge degree of freedom $k_j$, whereas in the spin sector,  the spin string states are given by 
\begin{eqnarray}
\label{string}
\lambda_{\alpha}^{n,j}=\lambda_{\alpha}^{n}+\frac{ic}{2}(n+1-2j), \quad j=1,2,\cdots,n,
\end{eqnarray}
which are called the length-$n$ spin strings. 
Such a spin structure comprises a rich magnetism like what the 1D Heisenberg spin chain has. 
Accordingly, the  Bethe ansatz equations (\ref{BA1}) and (\ref{BA2}) with string hypothesis (\ref{string}) reduce to the following two sets of  BA equations associated with the quantum number $\{I_j\}$ and $\{J_{\alpha}^n \}$
\begin{eqnarray}
&&k_jL=2\pi I_j - \sum_{n=1}^{\infty} \sum_{\alpha=1}^{M_n} \theta \left( \frac{2(k_j-\lambda_{\alpha}^n)}{nc} \right),  \label{BAE1-s}\\
&&\sum_{j=1}^{N} \theta \left( \frac{2(k_j-\lambda_{\alpha}^n)}{nc} \right) =2\pi J_{\alpha}^{n}\nonumber\\
&& + \sum_{m=1}^{\infty} \sum_{\beta=1}^{M_m} \Theta_{mn} \left( \frac{2(\lambda_{\alpha}^n-\lambda_{\beta}^m)}{c} \right),\label{BAE2-s}\
\end{eqnarray}
where  $j=1,2,\cdots,N$, $ \alpha =1,2,\cdots,M_n, \; n\geq 1$ and  $M_n$ is the number of length-$n$ string, $\theta(x)=2 \tan^{-1} (x)$, and the function $\Theta_{mn}(x)$ is defined by
\begin{eqnarray}
\Theta_{mn}(x) =  \left \{
\begin{array}{rcl}
&&\theta\left( \frac{x}{|n-m|} \right)+2\theta\left( \frac{x}{|n-m|+2} \right)+ \cdots \\
&&+2\theta\left( \frac{x}{n+m-2} \right)+\theta\left( \frac{x}{m+n} \right), \,\, \text{for}\,\,n \neq m,  \\
&&2\theta\left( \frac{x}{2} \right)+2\theta\left( \frac{x}{4} \right)+ \cdots \\
&&+2\theta\left( \frac{x}{2n-2} \right)+\theta\left( \frac{x}{2n} \right), \quad \text{for} \quad n = m.
\end{array}
\right.\nonumber
\end{eqnarray}
The quantum number $I_j$ for charge degree of freedom   takes  distinct integers (or half-odd integers) for even (odd) $\sum_\alpha  M_\alpha $, explicitly 
\begin{equation}
 I_j\in \sum_{n=1}^{\infty} \frac{M_n}{2} + \mathbb{Z}.
\end{equation}
The spin quantum number $J_{\alpha}^n$ are distinct integers (half-odd integers) for odd (even) $N-M_m$, which satisfy
\begin{eqnarray}
&& J_{\alpha}^n \in \frac{N-M_n}{2}+\frac{1}{2} + \mathbb{Z}, \nonumber \\
&&|J_{\alpha}^n | \leq J_{+}^n= \frac{N}{2} -\sum_{m=1}^{n}mM_m - n \sum_{m=n+1}^{\infty} M_m +\frac{M_n}{2} -\frac{1}{2}, \nonumber \\
&& J_{\alpha}^n = -J_{+}^n, -J_{+}^n+1, -J_{+}^n+2,\cdots, J_{+}^n-1,J_{+}^n.\nonumber \nonumber
\end{eqnarray}
The total momentum of the system is given  by
\begin{equation}
\label{totalmomentum}
K=\sum_{j=1}^{N}k_j = \frac{2\pi}{L} \left( \sum_{j=1}^{N} I_j + \sum_{\alpha=1}^{M_n} \sum_{n=1}^{\infty} J_{\alpha}^n  \right).
\end{equation}

\begin{figure}[tbp]
	\centering
	{\includegraphics[width=3.4in]{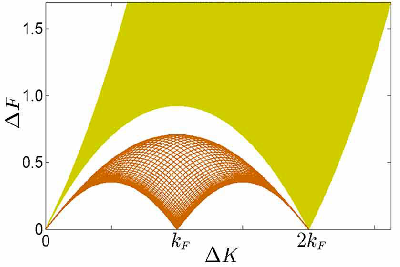}} 	
	\caption{{\small Exact low energy excitation spectra for charge (yellow green) and spin (dark brown) at $\gamma = 5.03$ with the Fermi surface $k_F=n\pi$, density $n=N/L=3 \times 10^6 $. The yellow green  spectrum shows the particle-hole continuum excitation. The dark brown  spectrum shows the notable  two-spinon excitation spectra. In long wave limit, i.e. $\Delta K\ll 1$, the spin and charge spectra show two independent linear dispersion with  velocities $v_s$ and  $v_c$, respectively. From He et. al. \cite{He:2020}. 
	 } }
	\label{spectrum}
\end{figure}

It is significantly found \cite{He:2020} from the above Bethe ansatz equations  (\ref{BA1}) and (\ref{BA2})  that the excitations in charge sector display  a particle-hole continuum spectrum, see Fig.~\ref{spectrum}. 
It shows a linear dispersion structure for arbitrary strongly interacting fermions in a  long wave limit
\begin{equation}
\omega (q)=v_c |q| \pm \frac{\hbar q^2}{2 m^*}+\cdots, \label{dispersion} 
\end{equation}
where the charge velocity and the effective mass are given by the following expressions 
\begin{eqnarray}
   v_c &=& \frac{\varepsilon_c'(k_0)}{2\pi \rho_c(k_0)}, \\
	\frac{1}{2 m^*} &=&\frac{\varepsilon_c''(k_0)}{2(2 \pi \rho_c(k_0)^2} - \frac{\pi \rho_c'(k_0) \varepsilon_c'(k_0)}{(2 \pi \rho_c(k_0))^3},
\end{eqnarray}
respectively.
For a strong coupling limit,  the charge velocity and the  effective mass  are given by \cite{He:2020}
\begin{eqnarray}
\label{EMstrong}
v_c \approx 2 \pi n_c \left(  1 - \frac{4 \ln 2}{\gamma} \right), \quad m^*=m\left(1+ \frac{4\ln2}{\gamma}\right).
\end{eqnarray}

The low-energy spin flipping excitation in the spin sector is also displayed in Fig.~\ref{spectrum}.
 It shows a typical two-spinon excitation spectrum in spin elementary excitations for the Yang-Gaudin model. 
 The total excited momentum is given by 
\begin{equation}
\label{spinexcitationmomentum}
\Delta K_{\rm spinon} =n\pi- 2\pi \sum_{j=1}^{2} \int_{0}^{\lambda_j^{\rm h}} \rho_s^0(\lambda) {\rm d}\lambda,
\end{equation}
while the energy of two-spinon excitation is given by 
\begin{align}
\label{spinexcitationenergy}
\Delta E_{\rm spinon} =-\sum_{j=1}^{2}\phi_s^0 (\lambda_j^{\rm h}),
\end{align}
that  present a  microscopic origin of the two deconfined spinons with spin-$1/2$, showing a fractional excitation. 
This result confirms  an antiferromagnetic ordering in the Fermi gas with  internal degrees of freedom. 
However, when the interaction increases, the spin excitation band becomes lower, and even it vanishes in the limit $\gamma \to \infty$. 

Moreover, Fig.~\ref{spectrum} remarkably displays the origin of the separated excitations in spin and charge  sectors. 
In fact, in one dimension such low-lying excitations with different sound velocities form two collective motions of bosons, i.e. the  so-called spin-charge separation -- fermions dissolve  into  spinons and holons. 
The spin-charge separation phenomenon is the  hallmark of one-dimensional  physics and has not been unambiguously confirmed by experiments, either in solids \cite{Kim:1996,auslaender2005spin,Kim:2006,Jompol:2009} or ultracold atoms \cite{Hulet:2018,Vijayan:2020}.  
For theoretical understanding such unique 1D many-body phenomenon, it acquires low temperature thermodynamics and dynamical correlation functions for  such excitations in spin and charge degrees of freedom. 
Such novel physics spin-charge separation has been recently observed with ultracold atoms in  \cite{Senaratne:2021}

On the other hand, the finite temperature problem for the Lieb-Liniger Bose gas was solved by C. N. Yang and C. P. Yang in 1969 \cite{Yang-Yang}.
Extension to the Yang-Gaudin model in terms of the Bethe ansatz equations (\ref{BAE1-s}) and (\ref{BAE2-s}) was done some time ago by Lai \cite{Lai:1971,Lai:1973} and  M Takahashi \cite{Takahashi:1971}, namely, the so-called thermodynamic Bethe ansatz (TBA) equations for the Yang-Gaudin model  are given by 
\begin{eqnarray}
	\varepsilon (k) &=& k^2-\mu-\frac{H}{2}-T \sum_{n=1 }^{\infty}a_n*{\rm ln} [1+{\rm e}^{- \phi_n (\lambda)/T}], \label{wholeTBA1}\quad \\
	\phi_n (\lambda)&=& nH-T a_n*\ln [1+{\rm e}^{- \varepsilon (k)/T}]\nonumber \\
	&+& T\sum_{m=1}^{\infty} T_{mn}*{\rm ln} [1+{\rm e}^{- \phi_m (\lambda)/T}] \label{wholeTBA2}
\end{eqnarray}
with $n=1,\ldots, \infty$. 
Here $*$ denotes the convolution, $\varepsilon(k)$ and $\phi_n(\lambda)$ are the dressed energies for the charge and the length-$n$ spin strings, respectively, with $k$'s and $\lambda$'s being the rapidities; and the function $T_{mn}(x) =d \Theta_{mn}(x) /d x$ is  given in Refs.~\cite{Takahashi:1971}. 
The pressure is given by
\begin{equation}
p=\frac{T}{2 \pi}\int_{-\infty}^{\infty} {\rm ln} [1+{\rm e}^{- \varepsilon (k)/T}] {\rm d}k, 
\end{equation}
from which all the thermal and magnetic quantities can be derived according to the standard statistical relations.

\begin{figure}[t]
	\begin{center}
	\includegraphics[width=1.0\linewidth]{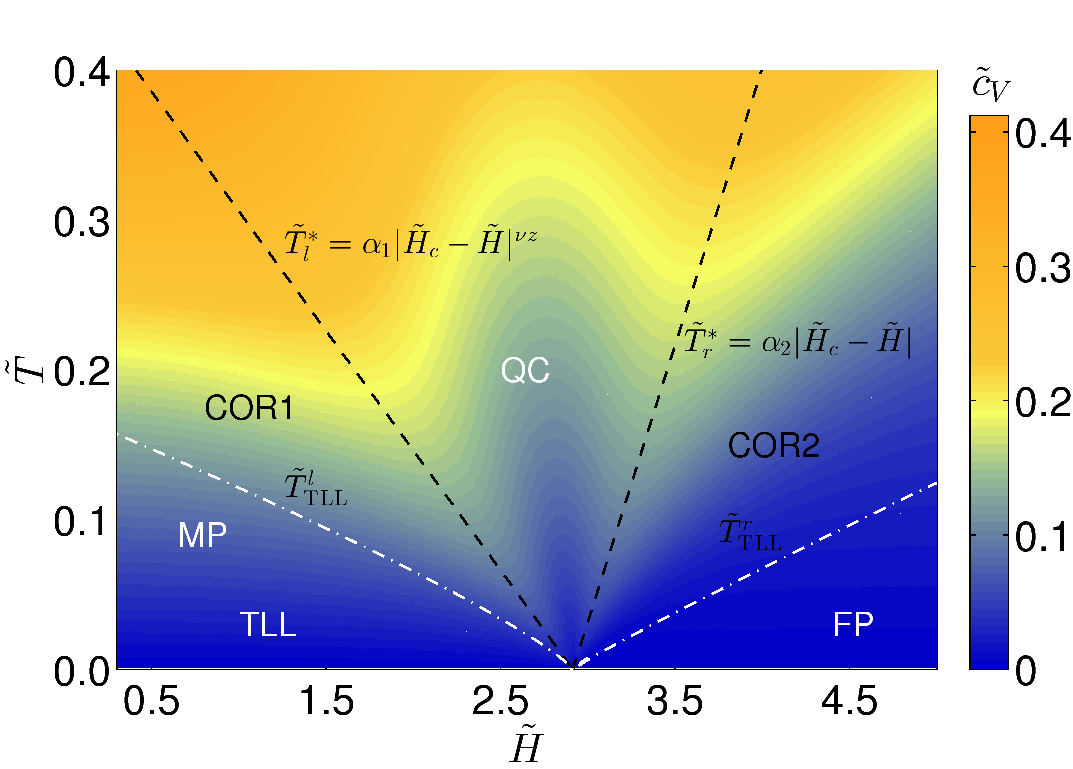}
	\end{center}
	\caption{(color online) Contour plot of  the dimensionless specific heat $\tilde{c}_V=c_v/T$, showing  the phase diagram in the $\tilde{T}$-$\tilde{H}$ plane. Here  dimensionless chemical potentials  $\tilde{\mu}=2.5$, $\tilde{H}_c=2.9145$. The black dashed lines denote the peak positions of specific heat, whereas  the white-dot-dashed line shows the boundary of the linear $T$ dependence of specific heat. $COR1$ denotes the crossover regions between QC and the TLL, giving a spin incoherent liquid. From He et. al. \cite{He:2020}. }
	\label{Fig:phasediagram}
\end{figure}

At low temperatures, the TBA equations (\ref{wholeTBA1}) and  (\ref{wholeTBA2}) are extremely hard to be solved either numerically or analytically. However, one can solve these coupled integral equations for certain physical regimes, for examples $k_BT \ll \mu, H$ or $c ^2 \gg E_F$, etc. 
For a fixed chemical potential,  the contour plot of the specific heat in  temperature-magnetic-field plane  naturally reads off different critical regions near the quantum phase transition from the  spin charge separated Tomonaga-Luttinger liquid (TLL) phase to a fully-polarized phase (FP), see Fig.~\ref{Fig:phasediagram}.
The quantum critical region fans out from the critical point, forming a critical cone. 

For $T\ll E_F$, we can safely neglect the contributions from the high strings  and just retain the leading length-1 string in the TBA equations (\ref{wholeTBA1}) and  (\ref{wholeTBA2}). 
Throughout the TLL phase with $H< H_c$, where $H_c$ is the critical field for a fixed chemical potential  
\begin{equation}
H_c=\left(\frac{c^2}{2 \pi }+2 \pi  n^2\right) \tan^{-1}\left(\frac{2 \pi n}{c}\right)-c n.
\end{equation}
The pressure can, in general, be   given by
\begin{equation}
p-p_0=\frac{\pi T^2}{6} \left(  \frac{1}{v_c}+\frac{1}{v_s} \right), \label{SC-Separation}
\end{equation}
where $p_0=\int_{-k_0}^{k_0}\varepsilon (k) dk$ is the pressure at $T=0$ and  the charge and spin velocities are given by 
\begin{equation} 
\label{vcvsdefination}
v_c=\frac{t_c}{2 \pi \rho(k_0)}\,,\;\;\;\; v_s=\frac{t_s}{2 \pi \sigma_1(\lambda_0)}\,,
\end{equation}
respectively,  with $\rho$ and $\sigma_1$ being the distribution functions  at  Fermi points $k_0$ and $\lambda_0$  for the charge and the spin sectors, respectively.  
While  $t_c=\frac{{\rm d}\varepsilon(k) }{{\rm d} k} \bigg|_{k=k_0}$ and $t_s=\frac{{\rm d}\phi_1(\lambda) }{{\rm d} \lambda} \bigg|_{\lambda=\lambda_0},$ are the respective linear slopes of the dispersion at the Fermi points.  
Although the proof of the form (\ref{SC-Separation}) is rather cumbersome, it gives a simple universal low temperature form of spin-charge separation theory \cite{Affleck1986} . The velocities can be numerically obtained  for arbitrary interaction strength. For the strongly interacting regime,  the charge and spin excitation velocities are given by 
\begin{equation}
v_c\approx 2\pi n\left(1-\frac{4\ln2}\gamma \right), \qquad
v_s\approx\frac{2\pi^3 n}{3\gamma}\left(1-\frac{6\ln
2}{\gamma}\right), \nonumber
\end{equation}
respectively. 
For the external field approaching the saturation field $H_c$,  the charge and spin velocities can be derived  from the
relations (\ref{vcvsdefination}).
The leading terms in the velocities are then found to be
\begin{eqnarray}
v_{c}&=&2\pi
n\left(1-\frac{12}{\pi\gamma}\sqrt{1-\frac{H}{H_{c}}}\right),\,\,
v_{s}=\frac{H_{c}}{n}\sqrt{1-\frac{H}{H_{c}}}.\nonumber
\end{eqnarray}

In the TLL regime,  
the dispersion relations for Yang-Gaudin model  are approximately
linear. 
Conformal field theory  predicts that the energy per unit
length has a universal finite size scaling form $E=E_{0}+\Delta/L^2$
where $E_{0}$ is the ground state energy per unit length for the
infinite system and $\Delta$ is a universal term. 
In this scenario,  Cardy \cite{Cardy1986}
showed that the two-point correlation function between primary
fields can be directly derived from conformal mapping using transfer
matrix techniques and expressed the conformal dimensions in terms of
finite-size corrections to the energy spectrum. 
For example, at zero temperature and zero magnetic field, by using conformal field theory \cite{Belavin1984,Blote1986}, the asymptotic of single particle correlation function can be given explicitly  
\begin{eqnarray}
&& G_{\uparrow}(x,t) \sim 
\langle\psi_{\uparrow}^{\dagger}(x,t)\psi_{\uparrow}(0,0)\rangle\nonumber\\
&&
\approx \frac{A_{\uparrow,1} e^{-\mathrm{i} k_{F\uparrow} x} }
{(x-\mathrm{i}v_{c}t)^{2\Delta _{c}^+}(x+\mathrm{i}v_{c}t)^{2\Delta_{c}^-}(x-\mathrm{i}v_{s}t)^{2\Delta_{s}^+}}
+h.c.
\end{eqnarray}
with the conformal dimensions 
\begin{eqnarray}
 \nonumber
2\Delta_{c}^{+}&=&\frac{9}{16}-\frac{3\ln
2}{4\gamma}+\frac{3}{2\pi^{2}}\left(\frac{H}{H_{c}}\right)+\frac{3}{2\gamma}\left(\frac{H}{H_{c}}\right)\nonumber\\
&&
-\frac{\ln 2}{\pi^{2}\gamma}\left(\frac{H}{H_{c}}\right), \label{CFD1}\\
 2\Delta_{c}^{-}&=&\frac{1}{16}-\frac{3\ln
2}{4\gamma}-\frac{1}{2\pi^{2}}\left(\frac{H}{H_{c}}\right)+\frac{1}{2\gamma}\left(\frac{H}{H_{c}}\right)\nonumber\\
&&+\frac{3\ln 2}{\pi^{2}\gamma}\left(\frac{H}{H_{c}}\right), \label{CFD2}
\\ 
2\Delta_{s}^{+}&=&\frac{1}{2}-\frac{2}{\pi^{2}}\left(\frac{H}{H_{c}}\right)-\frac{1}{\gamma}\left(\frac{H}{H_{c}}\right)
+\frac{4\ln 2}{\pi^{2}\gamma}\left(\frac{H}{H_{c}}\right). \label{CFD3}
\end{eqnarray}
We see clearly that the correlation function decays as some
power  of distance governed by the critical exponent.

On the other hand, universal scaling behaviour can be derived in the vicinity of the critical point at low temperatures. 
Near the critical point, the spin dressed energy $\phi_1(\lambda)$  in (\ref{wholeTBA2}) only has  a small negative part, which makes a major contribution to  the spin dressed energy near the critical point at low temperatures. 
Therefore, we can expand  the integration  kernel $a_n(k-\lambda)$ in terms of the  functions of small variables $\lambda$. After a tedious calculation, 
we obtain the pressure of  the Yang-Gaudin model  (\ref{Ham}) near the phase transition from the TLL phase to FP phase
\begin{eqnarray}
p &\approx& p_0-  \frac{ \arctan \left(\frac{2}{c} k_0 \right) T^{3/2} }{ \pi^{3/2} \sqrt{ (a+c_2)}}  \tmop{Li}_{\frac{3}{2}} \left( -{\rm e}^{\frac{s_0 \Delta H -c_1}{T}} \right) \nonumber\\
&&+\frac{ T^{5/2}}{4 \pi^{3/2} (a+c_2)^{3/2} } \frac{c k_0}{ \left( c^2/4+k_T^2 \right)^2}  \tmop{Li}_{\frac{5}{2}} \left( -{\rm e}^{\frac{s_0 \Delta H -c_1}{T}} \right), \label{pressure-QC} 
\end{eqnarray}
where the  regular part  $p_0$ of the pressure  is given by 
\begin{eqnarray}
\label{backgoudprssure}
p_0 = \frac{\pi T^2}{6 v_c}+ \frac{2}{3 \pi} \left( \mu_c +\frac{H}{2} \right)^{3/2} = p_0^{\rm Liquid} +p_0^{\rm BG}.
\end{eqnarray}
The above expression  $p_0^{\rm BG}=\frac{2}{3 \pi} \left( \mu_c +\frac{H}{2} \right)^{3/2}$ can be regarded as the background part of charge, whereas $  p_0^{\rm Liquid}=\frac{\pi T^2}{6 v_c}$ 
denotes the Luttinger liquid contribution from charge degrees of freedom. 
However,  the Luttinger liquid in the spin sector dissolves into the free fermion quantum criticality in the quantum critical (QC) region, see Fig.~\ref{Fig:phasediagram}.
It is worth noting that in the QC region the pressure (\ref{pressure-QC}) presents   a universal scaling form of the  equation of states
\begin{eqnarray}
p=p_0^{\rm Liquid}+p_0^{\rm BG}+T^{\frac{1}{z}+1} \mathcal{G}\left( \frac{s_0 \Delta H}{T^{1/\nu z}} \right), 
\end{eqnarray} 
from which one reads off the dynamical critical exponent $z=2$ and correlation length critical exponent $\nu=1/2$. 
Consequently, the scaling functions of  all  thermodynamic quantities can be derived based on this exact expression of the equation of states. 
The two crossover temperatures 
of the QC region  are given by $T_{l}^{*} =\alpha_1 |H-H_c|^{\nu z}$ and $T_{r}^{*} =\alpha_2 |H-H_c|$, here $\alpha_{1,2}=s_0/y_{1,2}$ with $y_1=-1.5629$, $y_2=3.6205 $ are numerical constants.
In the QC region, all thermodynamic quantities can be cast into universal scaling forms. 

In Fig.~\ref{Fig:phasediagram}, we also  identify a crossover region COR1 in the  region $E_{\rm spin}\ll k_BT \ll E_{F}$. Here $E_{\rm spin}$ and $E_F$ are the energy of spin sector and Fermi energy, respectively. 
In this region, the interplay between the spin and the charge degrees of freedom leads to a large deviation from the linear dispersion in  the spin  sector, leading to  a large  disruption of the spin charge coherent TLLs.  
Consequently, the spin-spin correlation function exponentially decays, while the charge-charge correlation still remains power law decay with distance. 
Breakdown of the  conformal field theory gives  rise to an  asymptotic of single particle correlation function 
\begin{eqnarray}
G_{\uparrow}(x,t) 
&\approx & \frac{A_{\uparrow,1} e^{-\mathrm{i} k_{F\uparrow}x } }
{(x-\mathrm{i}v_{c}t)^{2\Delta _{c}^+}(x+\mathrm{i}v_{c}t)^{2\Delta_{c}^-} } \nonumber\\
&& \times \frac{(\pi T/v_s)^{2\Delta^+_s} }{  e^{\frac{2\pi T}{v_s} x{\Delta_{s}^+} } }
+h.c.
\end{eqnarray} 
The conformal dimensions can be  given by Eqs. (\ref{CFD1})-(\ref{CFD3}).
In this regime, the temperature is low enough in comparison with the Fermi energy, whereas it is high enough in comparison with the spin excitation energy. Therefore the spin velocity $v_s \to 0$ in this crossover region COR1, which is a reminiscence of the spin incoherent liquid \cite{Fiete:2007,Cheianov:2004}. 

The novel phase diagram of Fig.~\ref{Fig:phasediagram} reveals a key concept that the low-lying excitations near the Fermi points dissolve into two separate collective motions of charge and spin, i.e., TLLs of charge and spin. 
Evidence for the spin-charge separation was  reported in solid state materials.  but none of those existing experiments in this study  seemed  to provide a conclusive observation of the spin-charge separated Luttinger liquids. 
In a recent experiment with ultracold atoms presented a confirmative observation of this phenomenon through the spin and charge dynamic structure factors  \cite{Senaratne:2021}.

\section{III. FFLO Pairing Correlation and Universal Thermodynamics for the Attractive  Fermi Gas}

For the attractive regime, i.e. $c<0$, the root patterns of the BA equations (\ref{BA1}) and (\ref{BA2}) are significantly different from that of the model for  $c>0$. 
For an attractive interaction, the quasimomenta $\left\{ k_i\right\} $ of the fermions with different spins  form two-body bound states \cite{Takahashi-a,Gu-Yang},
i.e., $k_i=\lambda_i \pm   \mathrm{i} \frac{1}{2} c$, accompanied by the real spin parameter
 $\lambda_i$.  Here $i=1,\ldots,N_{\downarrow}$.  The excess fermions have real quasimomenta $\left\{ k_j\right\} $ with $j=1,\ldots, N-2N_{\downarrow}$.
Thus the Bethe ansatz equations for the ground state  are transformed into the Fredholm equations according  to  the densities of the pairs $\rho_2(k)$   and density of single fermionic  atoms $\rho_1(k)$.    They  satisfy the following Fredholm equations \cite{Yang,Takahashi-a}
\begin{eqnarray}
\rho_1(k)&=&\frac{1}{2\pi}+\int_{-Q_2}^{Q_2}a_1(k-\lambda  )\rho_2(\lambda )d\lambda,   \label{Fermi2-a1}\\
\rho_2(\lambda )&=&\frac{2}{2\pi}+\int_{-Q_1}^{Q_1}a_1(\lambda -k')\rho_1(k')dk'\nonumber\\
&&+\int_{-Q_2}^{Q_2}a_2(\lambda -\lambda ')\rho_2(\lambda ')d\lambda ' \label{Fermi2-a2}
\end{eqnarray}
with  the integration boundaries $Q_1$ and $Q_2$, which  are the Fermi points of the single particles and pairs, respectively.
Here  $Q_1$ and $Q_2$ can be determined  by the conditions 
\begin{eqnarray}
n&\equiv:& \frac{N}{L}= 2\int_{-Q_2}^{Q_2}\rho_2(k)dk+\int^{Q_1}_{-Q_1}\rho_1(k)dk,\nonumber \\
n_{\downarrow}&\equiv: & \frac{N_{\downarrow}}{L}=\int_{-Q_2}^{Q_2}\rho_2(k)dk.\label{density-a}
\end{eqnarray}
The ground state energy per length is given by
\begin{equation}
E=\int_{-Q_2}^{Q_2}\left(2k^2+E_B \right)\rho_2(k)dk+\int_{-Q_1}^{Q_1}k^2\rho_1(k)dk. \label{Fermi2-E-a}
\end{equation}
In the above equation, the binding energy is given by $E_B=c^2/2$. 

For an attractive interaction, there exist two Fermi seas in charge degree of freedom at the ground state. So that  the Fermi points $Q_1$ and $Q_2$ are finite. This allows us to asymptotically solve the BA equations (\ref{Fermi2-a1}) and  (\ref{Fermi2-a2}). For weak and strong attractions with an arbitrary polarization $P=(n_{\uparrow}-n_{\downarrow})/(n_{\uparrow}+n_{\downarrow})$, the ground state energy are respectively given by \cite{Guan:PRA12}
 \begin{eqnarray}
E&=&\frac{1}{3}n_{\uparrow}^3\pi^2+\frac{1}{3}n_{\downarrow}^3\pi^2-2|c|n_{\uparrow}n_{\downarrow}+O(c^2),  \label{e-r-wbc}\\
E&\approx&\frac{\hbar^2n^3}{2m}\left\{-\frac{(1-P)\gamma^2}{4}
+\frac{\pi^2(1-3P+3P^2+15P^3)}{48}\right.\nonumber\\
&&\left.+\frac{\pi^2(1-P)(1+P-5P^2+67P^3)}{48|\gamma|}\right. \nonumber\\
&&\left. +\frac{\pi^2(1-P)^2(1+5P+3P^2+247P^3)}{64\gamma^2}\right.\nonumber\\
&&\left.-\frac{\pi^2(1-P)}{1440|\gamma|^3}\left[-15+31125{P}^{4}+1861{\pi }^{2}{P}^{5}\right.\right.\nonumber\\
&&\left.\left. -15765P^5-659{\pi }^{2}{P}^{4} +346{\pi }^{2}{P}^{3}-14{\pi }^{2}{P}^{2} \right.\right.\nonumber\\
& &\left.\left.+\pi^2
P+{\pi }^{2}-105P-150P^2-15090P^3\right]\right\}, \label{Energy}
\end{eqnarray}
respectively. 
This  result was also obtained from dressed energy equations \cite{Guan2007,Wadati}.
We notice that the energy  (\ref{e-r-wbc}) continuously connects to the repulsive ground state energy (\ref{E-r-wb})  at $c\to 0$. But  this  does not mean that the energy analytically connects because of the divergence in the small region $c\to \mathrm{i} 0$, see a discussion \cite{Guan:FP,Takahashi:1970-m}. 

At finite temperatures, except the two-body bound states  $k_j=\lambda_j\pm   \mathrm{i} \frac{1}{2} c$ in charge sector, the 
 spin quasimomenta of the  BA equations (\ref{BA1}) and (\ref{BA2})  form complex strings
 $\lambda_{\alpha,j}^n=\lambda^n_{\alpha}+\mathrm{i} \frac12 (n+1-2j)c$ with
 $j=1,\ldots ,n$ \cite{Takahashi:1971,Guan2007}. 
 Here $\alpha=1,\ldots, N_n$ is the number of strings. 
 The equilibrium states are determined by the minimization condition of
the Gibbs free energy, which gives rise to a set of coupled nonlinear
integral equations.
In terms of the dressed energies $\epsilon_{2}(k) := T\ln( \rho_2^h(k)/\rho_2(k) )$ and
$\epsilon_{1}(k) := T\ln( \rho_1^h(k)/\rho_1 (k) )$ for paired and unpaired fermions \cite{Takahashi:1971,Guan2007,Schlottmann:1993}, the thermodynamic Bethe ansatz equations for the 1D attractive  Fermi gas are given by 
\begin{eqnarray}
\epsilon_{2}(k)&=&2(k^2-\mu-\frac14{c^2})+Ta_2*\ln(1+\mathrm{e}^{-\epsilon_{2}(k)/T} )
  \nonumber\\
& &+ \, Ta_1*\ln(1+\mathrm{e}^{-\epsilon_{1}(k)/{T}}),\nonumber\\
\epsilon_{
  1}(k)&=&k^2-\mu-\frac12{H}+Ta_1*\ln(1+\mathrm{e}^{-\epsilon_{2}(k)/{T}})\nonumber\\
& &-T\sum_{n=1}^{\infty}a_n*\ln(1+\eta_n^{-1}(k)),\nonumber\\
\ln
  \eta_n(\lambda)&=&\frac{nH}{T}+a_n*\ln(1+\mathrm{e}^{-\epsilon_{1}(\lambda)/{T}})\nonumber
\\&&+\sum_{n=1}^{\infty}T_{mn}*\ln(1+\eta^{-1}_n(\lambda)),\label{TBA-Full}
\end{eqnarray}
where $n=1,\ldots, \infty$ and the function $\eta_n(\lambda) := \xi^h(\lambda)/\xi(\lambda ) $ is the ratio of the
string densities.  While the function $T_{mn}(\lambda)$ is given in \cite{Takahashi:1971,Guan2007}.
The Gibbs free energy per unit length, i.e. the pressure,  is given by
\begin{eqnarray}
p&=&\frac{T}{\pi}\int_{-\infty}^{\infty}dk\ln(1+\mathrm{e}^{-\epsilon_{2}(k)/{T}})\nonumber\\
      &&+\,\frac{T}{2\pi}\int_{-\infty}^{\infty}dk \ln(1+\mathrm{e}^{-\epsilon_{1}(k)/{T}}).\label{pressure}
\end{eqnarray}
This serves as the equation of states, from which  all the thermal and magnetic quantities can be derived according to the standard statistical mechanic relations.

From the TBA equations (\ref{TBA-Full}), at low temperatures, the pressure of the attractive Fermi gas with a strong interaction  can be written as a sum of two components: $p=p_1+p_2$, where $p_1$ is the pressure for unpaired fermions and $p_2$ for pairs, given explicitly in~\cite{Guan:2011-QC}
\begin{equation}
 p_{1}= F^{1}_{\frac{3}{2}}\left[1
+\frac{p_2 }{4|c|^3} \right], \,p_{2}= F^{2}_{\frac{3}{2}}\left[1+\frac{4p_1}{|c|^3}
+\frac{p_2 }{4|c|^3}\right], \label{EOS-polylog}
\end{equation}
respectively. Here $F^{r}_{a}=-\sqrt{{r}/{4\pi}}T^a{\rm Li}_a[-\exp({A_{r}/T})]$ ($r=1,2$) with
$A_{1} = \mu_1-\frac{2p_2}{|c|} +
\frac{1}{4|c|^3}F^{2}_{\frac{5}{2}} +Te^{-\frac{H}{T}}e^{-\frac{J}{T}}I_0\left(\frac{J}{T}\right)$
and  $A_{2} = 2\mu_2-\frac{4p_1}{|c|}-\frac{p_2}{|c|}+
\frac{8}{|c|^3}F^{1}_{\frac{5}{2}}+
\frac{1}{4|c|^3}F^{2}_{\frac{5}{2}}$.
In the above equations $\mathrm{Li}_{n}(x)=\sum_{k=1}^\infty{x^k}/{k^n}$ is the polylogarithm function, $J=2p_{1}/|c|$ and $I_0(x)=\sum_{k=0}^{\infty}\left(x/2\right)^{2k}/{(k!)^2}$. 
This is a remarkable result of the low temperature thermodynamics of the Yang-Gaudin model \cite{Guan:2011-QC}.  Based on this result, the exact results of quantum criticality of the attractive Yang-Gaudin model can be found in \cite{Guan:2011-QC,Yin:2011,Chen:2014,He-WB:2016}.

\begin{figure}[htbp]
\includegraphics[width=1.0\linewidth]{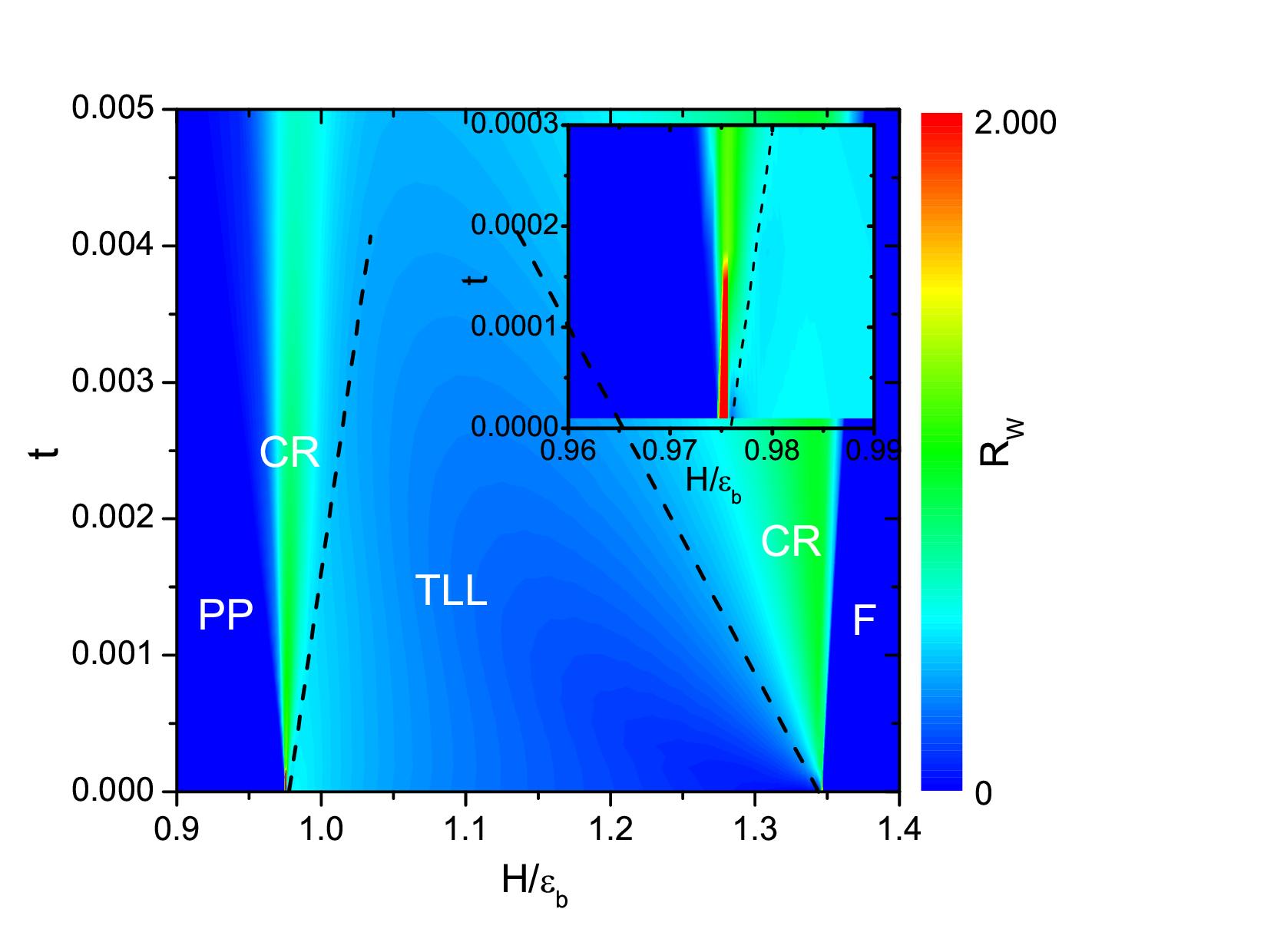}
\caption{(Color online) Contour plot of the susceptibility Wilson ratio $R_W^\chi$ (\ref{WR-sus}) of the attractive Fermi gas for dimensionless interaction $|\gamma| =10$. 
In this figure the reduced temperature $t=T/\varepsilon_b$ was used.  $\varepsilon_b$ denotes the binding energy. 
The susceptibility Wilson ratio is temperature independent in the FFLO-like phase
The dashed lines indicate the crossover  temperature $T^*=\alpha  |H-H_c|$ separating the relativistic TLL  liquid  from the free fermion quantum criticality. 
$R_W =0$ for both the TLL  of pairs (PP) and the TLL of excess fermions (F). From Guan {\em et. al.}, \cite{Guan:2013PRL}. }
\label{fig:contour}
\end{figure}

A significant feature of the Yang-Gaudin model with an attractive interaction is  the existence of the  FFLO pairing state \cite{Larkin1965,Fulde1964,Yang2001}.  The FFLO physics in Yang-Gaudin model  has  received extensive study in  theory \cite{Orso,HuiHu,Feiguin2007,Rizzi2008,Zhao2008,Lee2011,Schlottmann2012}  and experiment \cite{Liao,Revelle:2016}. 
 In 2007, three groups \cite{Guan2007,Orso,HuiHu}  predicted the phase diagram of the attractive Fermi gas where the novel FFLO-like pairing phase sits in a large parameter space.
 The phase diagram can be mapped out from specific heat or dimensionless ratios, such as Gr{\"u}neisen parameter \cite{Peng:2019,Yu:2020}, Wilson ratios \cite{Guan:2013PRL,Yu:2016}. 
The dimensionless susceptibility $\chi$ and compressibility $\kappa$  Wilson ratios are given by 
\begin{eqnarray}
R_W^{\chi} =\frac{4 }{3} \left( \frac{\pi k_B}{g \mu_B}\right)^2 \frac{\chi}{c_V/T}, \quad R_W^{\kappa} =\frac{\pi^2 }{3}  \frac{\kappa}{c_V/T}
 \label{WR-sus}
\end{eqnarray}
that map out the phase diagram of the model  in low temperature limit  \cite{Guan:2013PRL,Yu:2016}.

It is very interesting to note \cite{Guan:2013PRL,Yu:2016} that additivity rules of thermodynamical properties within subsystems of the FFLO-like phase are a reminiscence  of the rules for multi-resistor
networks in series and parallel.
Such simple additivity rules indicate  a novel and useful characteristic of multi-component TLLs   regardless  of microscopic details of the systems. 
For the $SU(2)$  Fermi gas (\ref{Ham}) with an attraction,  we have the relations $H/2 = \left( {\mu}_{ 1} -{\mu}_{ 2}\right) + E_B/2$ and  $n=2n_2+n_1$. Thus we may define the susceptibilities of subsystems in the FFLO  state
$\chi_{1} =(\mu_{\rm B} {g}_\mathrm{Lande} )^{2}  \left( {\partial n_{1}}/{\partial \mu_{1}}\right)_{n}$, 
$\chi_{2} =2(\mu_{\rm B} {g}_\mathrm{Lande} )^{2} \left( {\partial n_{2}}/{\partial \mu_{2}}\right)_{n}$, and the compressibility $\kappa_{ 1} = \left( {\partial n_{1}}/{\partial \mu_{1}} \right)_{H}$ and
$\kappa_{ 2} = 2\left( {\partial n_{2}}/{\partial \mu_{2}} \right)_{H}$ in the grand canonical ensemble.
Here $\mu_B$ is the Bohr magneton
and $g_\mathrm{Lande}$ 
is the Lande factor.
Consequently, the compressibility and susceptibility satisfy the following additivity rules 
\begin{eqnarray}
\kappa&=&\kappa_{ 1}+\kappa_{ 2},\label{compressibility}\\
\frac{1}{\chi}&=&\frac{1}{\chi_{1}}+\frac{1}{\chi_{2}}. \label{susceptibility}
\end{eqnarray}
The compressibilities and susceptibilities are given explicitly  in units of $\hbar^2 /(2m)$
\begin{eqnarray}
  \kappa_{\mathrm{2}}^{- 1} & \approx & \frac{\pi^2 n_2}{4}  \left( 1 +
  \frac{6 n_1}{|c|} + \frac{4 n_2}{|c|} + \frac{n_2^2}{2 |c|n_1} + \frac{24
  n_1^2}{c^2}  \right.
  \nonumber\\
  &  & \left.+ \frac{24 n_1 n_2}{c^2} + \frac{17 n_2^2}{c^2} - \frac{2 n_2^3}{c^2 n_1} + \frac{n_2^4}{4 c^2 n_1^2} \right),\label{com-ub}
  \\
  \kappa_{\mathrm{1}}^{- 1} & \approx & 2 \pi^2 n_1  \left( 1 + \frac{12
  n_2}{|c|} + \frac{16 n_1^2}{|c|n_2} + \frac{96 n_2^2}{c^2} + \frac{384
  n_1^2}{c^2}  \right.\nonumber \\
 && \left.- \frac{8 n_1 n_2}{c^2}- \frac{96 n_1^3}{c^2 n_2} + \frac{256
  n_1^4}{c^2 n_2} \right)
  \end{eqnarray}
  and 
  \begin{eqnarray}
  \chi_{1}^{-1}&=&  \pi^2 n_2\left[ 1+\frac{4}{|c|}(n-3n_2)+\frac{3}{c^2}(4n^2\right.\nonumber \\
  &&\left. -24nn_2+30n_2^2)\right],\\
  \chi_{2}^{-1}&=&  8\pi^2 n_1 \left[ 1+\frac{4}{|c|}(n-2n_1)+\frac{4}{c^2}(2n^2\right.\nonumber\\
  &&\left. +10n_1^2-12nn_1)   \right].
\end{eqnarray}

Moreover, the interaction effect is revealed from  the  sound velocities $v_{1}$, $v_{2}$  of the excess fermions and bound pairs, for example, for strong attraction 
\begin{eqnarray}
 \label{equ:sound_velecity_1}
v_{1}&\approx &\frac{\hbar }{2m}2\pi n_{1}   \left(1+ 8 n_{2} /|c| +48 n_{2}^2/c^2 \right),\nonumber \\
v_{2}&\approx &\frac{\hbar }{2m} \pi n_{2}  \left(1+ 2A/|c| +3A^2/c^2 \right),
\end{eqnarray}
with $A=2n_{1}+n_{2}$.
We similarly find that the specific heat satisfies 
\begin{equation}
c_{V}=c_{V,1}+c_{V,2}, \label{heat} 
\end{equation}
where $c_{V,r}=  \frac{\pi k_B^2T}{3 \hbar} \frac{1}{v_r}
$ with $r=1,\,2$. 
It is remarkably observed that such additivity rules of compressibility (\ref{compressibility}), susceptibility (\ref{susceptibility}) and  the rescaled specific heat $c_{V}/T$ with  (\ref{heat}) do not depend on the temperature in the FFLO-like state below a certain temperature. 
 Such additivity rules are a  characteristic of multi-component TLLs \cite{Yu:2016}.
This nature  presents a universal feature of thermodynamics for a two-component TLLs of pairs and single fermions in 1D. 

Moreover, it was proved  \cite{Yu:2016} that the susceptibility and compressibility  Wilson ratios
\begin{eqnarray}
   R^{\chi }_{\mathrm{W},r'}
&= &
\left( \sum_{r=1}^{2} \frac{D_{r}^{\chi}}{r^2}\right)^{-1}
\left( \sum_{r=1}^{2}\frac{1}{v_{r}} \right)^{-1},\label{WR1}\\
%
R^{\kappa}_\mathrm{W}
&= &
\left( \sum_{r=1}^{2} \frac{r^2}{D_{r}^{\kappa}} \right)
\left( \sum_{r=1}^{2} \frac{1}{v_{r}} \right)^{-1}. \label{WR2}
\end{eqnarray}
are dimensionless and uniquely determined by the sound velocities
and stiffnesses.
In the above expressions, the individual  stiffnesses $D^{\kappa}_r$ and $D^{\chi}_{r}$  can be  respectively given 
 in terms of $\kappa_r$ and $\chi_r$ via 
 \begin{eqnarray}
 D_{ r}^{\kappa}& = & \frac{r^2}{ \pi\hbar} \frac{1}{  \kappa_{r}},\qquad
D_{ r}^{\chi}=  \frac{r^2}{ \pi\hbar} \frac{1}{ \chi_{r} }.
\label{eqs}
\end{eqnarray}
In this context the Wilson ratios of the Yang-Gaudin model  with polarization elegantly  determine the 
TLL nature of the FFLO-like state, see 
Fig.~\ref{fig:contour}. 
It also  indicates a universal feature of the quantum criticality of the attractive Fermi gas.   

\begin{figure}[ht]
\begin{center} 
\includegraphics[width=0.5\textwidth]{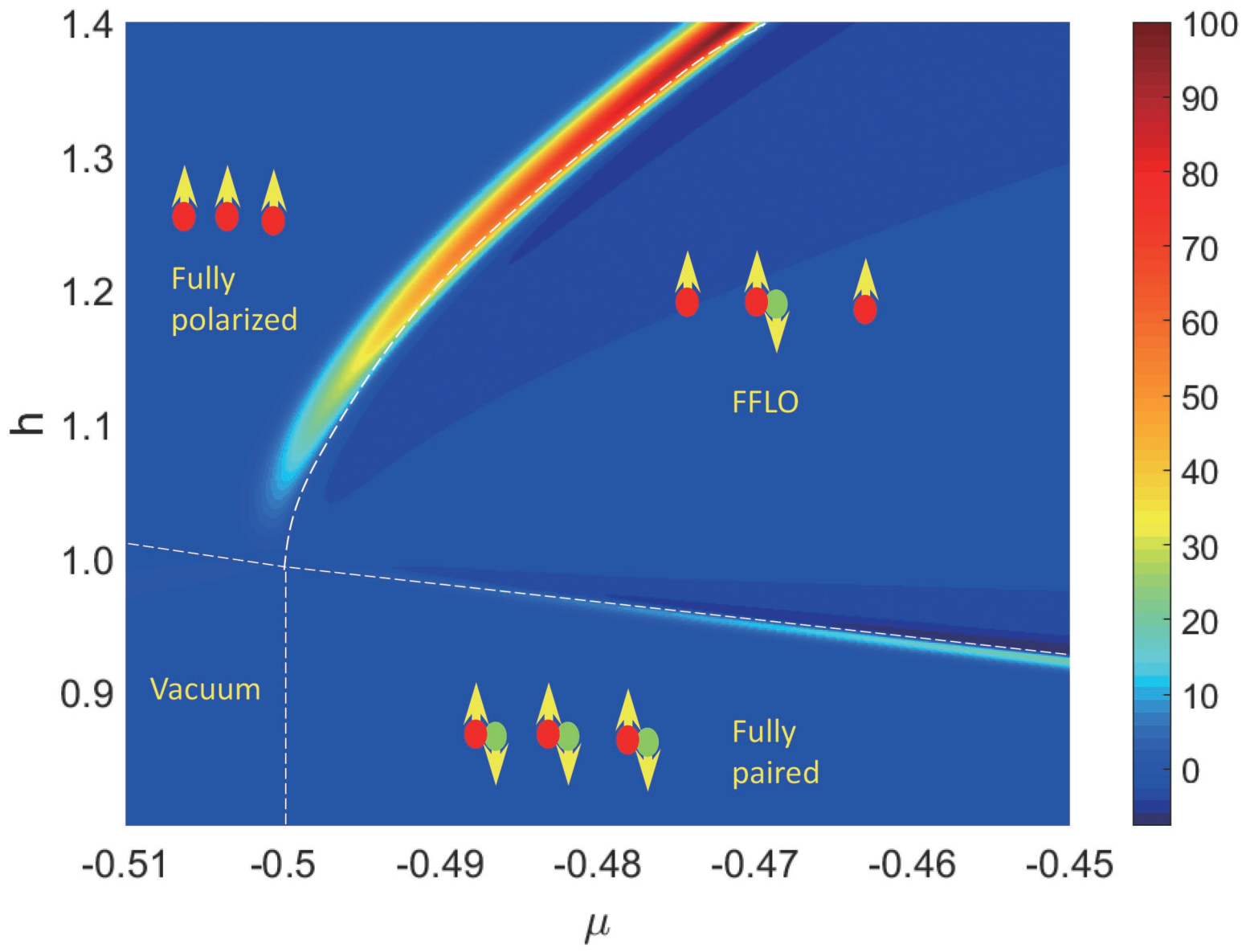}
\end{center}
\caption{Contour plot of the negative Gr\"uneisen parameter (\ref{eq:calculation_for_Gamma}), i.e. $-\Gamma$,  mapping out the full phase diagram of the Yang-Gaudin model with an attractive interaction in $h-\mu$ plane, see the main text.
Here we set  the dimensionless temperature  $t=0.001$. From Yu,  {\em et. al.} \cite{Yu:2020}. }
\label{phasediagram-a}
\end{figure}

On the other hand, Gr\"uneisen parameter \cite{Gruneisen_AdP_1908}, which was  introduced  in the beginning of 20th century in the study of the effect of volume change of a crystal lattice on its vibrational frequencies,  has been extensively studied for the exploration of the caloric effect of solids and phase transitions associated with volume change.
Recently, one of the authors of this paper and coworkers \cite{Peng:2019,Yu:2020}  studied interaction-  and chemical potential-driven caloric effects and Gr\"uneisen parameter in ultracold atomic gases in 1D. 
By using the Maxwell's relations , the Gr\"uneisen parameter of quantum gases  in grand canonical ensemble  is given by  \cite{Yu:2020} 
\begin{equation}
\Gamma=V\frac{\left.\frac{\mathrm{d}p}{\mathrm{d}T}\right|_{V,N}}{\left.\frac{\mathrm{d}E}{\mathrm{d}T}\right|_{V,N}}
=\frac{1}{T} 
\frac{\frac{\partial^2 p}{\partial \mu^2}\frac{\partial p}{\partial T}
-\frac{\partial^2 p}{\partial \mu \partial T}\frac{\partial p}{\partial \mu}}
{\frac{\partial^2 p}{\partial \mu^2}\frac{\partial^2 p}{\partial T^2}
-\left(\frac{\partial^2 p}{\partial \mu \partial T}\right)^2},
\label{eq:calculation_for_Gamma}
\end{equation} 
that remarkably maps out the phase diagram of the Yang-Gaudin model with an attractive interaction, see  Fig.~\ref{phasediagram-a}. 
This is intimately related to the expansionary caloric effect
\begin{eqnarray}
 \left.\frac{\partial T}{\partial V}\right|_{S,N,H}&=&\frac{T}{V} \Gamma. \label{GP-1}
\end{eqnarray}
Similarly one can find 
\begin{eqnarray}
\left.\frac{\partial T}{\partial H} \right|_{S,N,V}&=&\frac{T}{H} \Gamma_{\text{mag}},
\label{eq:MCE}\\
\left.\frac{\partial T}{\partial c} \right|_{S,N,V,H}&=&\frac{T}{c} \Gamma_{\text{int}}
\label{eq:CMCE}
\end{eqnarray}
that establish  important relations between magnetocaloric/interaction-driven caloric  effect and the Gr\"uneisen parameter, respectively.  
In the above equations, the magnetic and interaction Gr\"uneisen parameters in grand canonical ensemble  are given by 
\begin{eqnarray}
\Gamma_{\text{mag}}&=&-\frac{H}{T} \frac{\frac{\partial^2 p}{\partial \mu^2}\frac{\partial^2 p}{\partial H \partial T}
-\frac{\partial^2 p}{\partial \mu \partial H}\frac{\partial^2 p}{\partial \mu \partial T}}{\frac{\partial^2 p}{\partial \mu^2}\frac{\partial^2 p}{\partial T^2}
-(\frac{\partial^2 p}{\partial \mu \partial T})^2},
\label{eq:calculate_mag_gamma}\\
\Gamma_{\text{int}}&=&-\frac{\frac{\partial^2 p}{\partial \mu^2}\frac{\partial^2 p}{\partial c \partial T}
-\frac{\partial^2 p}{\partial \mu \partial c}\frac{\partial^2 p}{\partial \mu \partial T}}{\frac{\partial^2 p}{\partial \mu^2}\frac{\partial^2 p}{\partial T^2}
-(\frac{\partial^2 p}{\partial \mu \partial T})^2}
\frac{c}{T},
\end{eqnarray}
respectively.  It is interesting to note that like the adiabatic demagnetization cooling in solid,  the interaction ramp-up and -down in quantum gases  provides a promising protocol of quantum refrigeration.

The Yang-Gaudin model (\ref{Ham}) with an attractive interaction exhibits three phases of quantum states:
a fully paired phase with polarization $P=0$, a partially-polarized FFLO-like phase with $0<P<1$ as well as a fully-polarized phase with $P=1$ at zero temperature, see Fig.~\ref{phasediagram-a}. The  Gr\"uneisen parameter has a sudden enhancement near the quantum phase transition that  gives a universal divergent scaling $\Gamma
\sim  t^{-1/2} $. 
The key features of this phase diagram were experimentally confirmed using finite temperature density profiles of trapped fermionic ${}^6$Li atoms \cite{Liao}.

 In the FFLO-like phase, at zero temperature,  the leading order of the long distance asymptotics for the pair
correlation function $G_{p}(x,t)=\langle\psi_{\uparrow}^{\dagger}(x,t)\psi_{\downarrow}^{\dagger}(x,t)\psi_{\uparrow}(0,0)\psi_{\downarrow}(0,0)\rangle$ oscillates with wave number
$\pi(n_{\uparrow}-n_{\downarrow})\equiv\pi n_{f}P$ where
$n_{f}=n_{\uparrow}+n_{\downarrow}$. 
The proof can be straightforward by using the conformal field theory \cite{Lee2011}, i.e. 
\begin{eqnarray}
\nonumber G_{p}(x,t) &\approx&
\frac{A_{p,1}\cos\left(\pi(n_{\uparrow}-n_{\downarrow})x\right)}{|x+\mathrm{i}v_{u}t|^{\theta_{1}}|x+\mathrm{i}v_{b}t|^{\theta_{2}}},
\end{eqnarray}
where $\theta_{1} \approx \frac{1}{2}$ and $\theta_{2} \approx
\frac{1}{2}-\frac{(1-P)}{2|\gamma|}$ for a strong attraction.
 The
oscillations in $G_{p}(x,t)$ are caused by an
imbalance in the densities of spin-up and spin-down fermions, i.e.
$n_{\uparrow}-n_{\downarrow}$ which is a mismatch in Fermi
surfaces between both species. 
The spatial modulation is characteristic of the FFLO state. The backscattering among the Fermi points of bound pairs and unpaired fermions results in a 1D analog of the FFLO-like state and displays a microscopic origin of the FFLO pairing correlation.
These results are consistent with the
Larkin-Ovchinikov phase \cite{Larkin1965} and the wave numbers
coincide with the ones discovered through  the density matrix
renormalization group method
\cite{Feiguin2007,Tezuka2008,Rizzi2008}, quantum Monte Carlo method
\cite{Batrouni2008}, the mean field approach \cite{Liu2008,Zhao2008}
and bosonization technique \cite{Yang2001}.


\section{IV. Conclusion and Outlook}
We have presented  a brief review of recent developments of the  Yang-Gaudin model from integrability perspectives. 
It turns out that the Bethe ansatz solution of the model provides a rigorous understanding of many-body phenomena ranging from fractional excitations to  spin charge separation, FFLO pairing state and universal thermodynamics  and  quantum criticality as well. 
The legacy of Yang-Baxter equation significantly   contributes to  developments of analytical methods for cold atoms, spin liquids and condensed matter physics, particularly with regard to the fundamental many-body physics to be gained from exactly solvable quantum many-body systems. In this short review, we have also discussed  a number of the authors' contributions in the study of the Yang-Gaudin model, which have led to direct applications in recent breakthrough experiments on low-dimensional many-body physics of ultracold atoms \cite{Liao,Yang:2017,Hulet:2018,Revelle:2016,Senaratne:2021}. An outlook for future research  on  the Yang-Gaudin  model includes: 

(a) The observation of spin-charge separation phenomenon is  a promising research in many-body physics.  
 People,  in several important  papers  \cite{Kim:1996,auslaender2005spin,Kim:2006,Jompol:2009,Hulet:2018,Vijayan:2020},   experimentally probed spin-charge separation with evidence.
Recent new experiment \cite{Senaratne:2021} has  provided a conclusive observation of the spin-charge separated Luttinger liquids.  This experimental confirmation of this 1D unique phenomenon  should  include: 1) identification of the separate collective excitation spectra of charge and spin, 2) confirmation of the spin and charge dynamical response  correlation functions; 3) determination of the independent spin and charge sound velocities and their Luttinger parameters. This opens to further study of spin coherent and incoherent Luttinger liquids in quantum gases with higher spin symmetries.  Applications of such unique 1D  behaviour in quantum metrology and quantum information will be highly expected.

(b) The wave function of the Yang-Gaudin model is largely unexplored due to the complexity of the Bethe's  superpositions of $N!$ plane waves. So far there has been a little understanding of correlation functions and dynamical response functions for ground state and excited states of this model. Such a lack of  study prevent an access to quantum entanglement behaviour and  metrological useful information for quantum technology. It was recently shown \cite{Hauke:2016} that  the dynamical response function can be used to  measure multipartite entanglement in quantum  spin systems.  This opens a promising  opportunity to further  explore  realistic applications of fractional excitations,  spin liquids  and impurities   in quantum metrology. 

(c) Cooling fermions in ultracold atomic experiments  remain elusive. To achieve this goal, it is essential to understand caloric effects induced by magnetic field, trapping potential and dynamical interaction in quantum gases. 
Therefor in this research there exist  many open questions regarding adiabatic processes and heat exchanges between the system and baths.  The Yang-Gaudin model has rich phases of quantum matter which hold a promise for studying   quantum transport, hydrodynamics, quantum heat engine and quantum refrigeration by driving external trapping potentials and interactions. 

 {\bf Acknowledgement}  This article particularly delicates to the centenary of Professor C. N. Yang's birthday.
 XWG is very  grateful to professor C N Yang for his mentoring, constant help and encouragements since the first time I met him in 2010. He also acknowledges Institute for Advanced study,  Tsinghua University  and Beijing Computational Science Research Center for their kind hospitality. 
 X.-W. G. is supported by NSFC Key Grant No.12134015 and NSFC Grant No.11874393. 
 HQL thanks
professor C N Yang for his endless support and great
contributions to the physics department at the Chinese
University of Hong Kong.


\begin{thebibliography}{99}

\bibitem{Bethe:1931} H. Bethe,  {\em Z. Physik},  {\bf 71},  205 (1931).


 \bibitem{Hulthen} L. Hulthen\'{e}n, Arkiv. Mat. Mstron. Fysik  {26 A}, 11 (1938).
 
  \bibitem{Orbach} R. Orbach, Phys. Rev. {\bf 112}, 309 (1958).
  
   \bibitem{Walker} L. R. Walker, Phys. Rev. {\bf 116}, 1089 (1959). 
 
 \bibitem{Griffith} R. B. Griffiths, Phys. Rev. {\bf 113}, A768 (1964).

\bibitem{Cloizeau}J. des Cloizeaux and J. J. Pearson, Phys. Rev. {\bf 128}, 2131 (1962). 

\bibitem{YY-1}C. N. Yang and C. P. Yang, Phys. Rev. {\bf 150}, 321 (1966);

\bibitem{YY-2}  C. N. Yang and C. P. Yang, Phys. Rev. {\bf 150}, 327 (1966);
 
\bibitem{YY-3}C. N. Yang and C. P. Yang, Phys. Rev. {\bf 151}, 258 (1966).

\bibitem{Lieb-Liniger}E. H. Lieb and W. Liniger,  { Phys. Rev.} {\bf 130}, 1605 (1963).

\bibitem{Yang}C. N. Yang, { Phys. Rev. Lett.} {\bf 19},  1312 (1967).

\bibitem{Gaudin}M. Gaudin,  { Phys. Lett. A} {\bf 24},  55 (1967).

\bibitem{Baxter} R. J. Baxter, Ann. Phys. (N. Y.) {\bf 70}, 193 (1972);
\newline 
R. J. Baxter, Ann. Phys. (N. Y.) {\bf 70}, 323 (1972).



\bibitem{Lieb-Wu}E. H. Lieb and F. Y.  Wu, Phys. Rev. Lett. {\bf  20}, 1445 (1968).

\bibitem{Sutherland:1968} B. Sutherland, Phys. Rev. Lett. {\bf  20}, 98 (1968).

\bibitem{Andrei:1983}N. Andrei, K. Furuya, and J. H. Lowenstein, Rev. Mod. Phys.
{\bf 55}, 331 (1983).

\bibitem{Dukelsky:2004}J. Dukelsky, S. Pittel, and G. Sierra,  Rev. Mod. Phys. {\bf 76}, 643 (2004).


\bibitem{1D-Hubbard}Essler F H L, Frahm H, Göhmann F, Klümper A and Korepin V E 2005 {\em The One-Dimensional Hubbard Model} (Cambridge: Cambridge University Press)



\bibitem{Korepin}V. E. Korepin,  A. G. Izergin, and N. M.  Bogoliubov,  { Quantum Inverse Scattering Method and
Correlation Functions} (Cambridge: Cambridge University Press), 1993.

\bibitem{Sutherland-book}B. Sutherland, { Beautiful Models: 70 years of exactly solved quantum many-body problems} (Singapore: World Scientific), 2004.

\bibitem{Takahashi-b}M. Takahashi { Thermodynamics of One-Dimensional Solvable Models} (Cambridge: Cambridge University Press), 1999.

\bibitem{Wang-book} Y.-P. Wang, W.-L. Yang, J. Cao and K.-J. Shi, {Off-Diagonal Bethe Ansatz for Exactly Solvable Models}, (Springer-Verlag Berlin Heidelberg), 2015.

\bibitem{Batchelor:2007}M. T. Batchelor, X. W. Guan, N. Oelkers, and Z. Tsuboi, Adv. Phys. {\bf 56}, 465 (2007).


\bibitem{Cazalilla:2011}M. A. Cazalilla, R.  Citro, T. Giamarchi,  E. Orignac  and M. Rigol,  { Rev. Mod. Phys.} {\bf 83} 1405 (2011).

\bibitem{yangyou} C. N. Yang and Y. Z. You, Chin. Phys. Lett. {\bf 28}, 020503 (2011).  

\bibitem{Guan:2013}X. W. Guan, M. T. Batchelor and C. Lee,  { Rev. Mod. Phys.} {\bf 85},  1633 (2013). 

\bibitem{Batchelor:2016} M. T. Batchelor and  A. Foerster, J. Phys. A: Math. Theor. {\bf 49}, 173001 (2016).

\bibitem{Mistakidis:2022} S. I. Mistakidis, et. al. arXiv:2202.11071

\bibitem{Yang-Yang}C. N. Yang  and C. P. Yang, {J. Math. Phys.} {\bf 10},  1115 (1969).  


\bibitem{Takahashi:1971}M. Takahashi,  Prog. Theor. Phys. {\bf 46}, 401 (1971);
\newline
M. Takahashi, Prog. Theor. Phys. {\bf 46}, 1388 (1971).
\bibitem{Takahashi:1972}M. Takahashi, Prog. Theor. Phys. {\bf 47}, 69 (1972);
\newline 
M. Takahashi,   Prog. Theor. Phys. {\bf 50}, 1519 (1973);
\newline
M. Takahashi,  Prog. Theor. Phys. {\bf 52}, 103 (1973).



\bibitem{Guan2007} X. W. Guan, M. T. Batchelor, C. Lee and M. Bortz,
Phys. Rev. B {\bf 76}, 085120 (2007)


\bibitem{Zhao} E. Zhao, X.-W. Guan, W. V. Liu, M. T. Batchelor and M. Oshikawa, Phys. Rev. Lett. \textbf{103}, 140404 (2009).

\bibitem{Guan:2013PRL}
X.-W. Guan, X.-G. Yin, A.~Foerster, M.~T. Batchelor, C.-H. Lee, and H.-Q. Lin,
Phys. Rev. Lett., {\bf 111}, 130401 (2013).


\bibitem{Yang:2017}
Bing Yang, Yang-Yang Chen, Yong-Guang Zheng, Hui Sun, Han-Ning Dai, Xi-Wen
  Guan, Zhen-Sheng Yuan, and Jian-Wei Pan, 
 Physical review letters,  {\bf 119}, 165701 (2017).

\bibitem{He:2020}Feng He, Yu-Zhu Jiang, Hai-Qing Lin, Randall G. Hulet,  Han Pu,  and Xi-Wen Guan, 
 Physical review letters,  {\bf 125}, 190401 (2020).


\bibitem{PhysRevB.101.035149}
Ovidiu~I. P\^a\ifmmode~\mbox{\c{t}}\else \c{t}\fi{}u, Andreas Kl\"umper, and
  Angela Foerster.
Physical Review B, {\bf 101}, 035149 (2020).


\bibitem{Olshanii_PRL_1998}
M.~Olshanii, 
Physical Review Letters, {\bf 81}, 938 (1998).

\bibitem{Guan:PRA12} X.-W. Guan, Z.-Q. Ma, Phys. Rev. A {\bf 85}, 033632 (2012).


\bibitem{Kim:1996}
C~Kim, AY~Matsuura, Z-X Shen, N~Motoyama, H~Eisaki, S~Uchida, Takami Tohyama,
  and S~Maekawa,
 Physical review letters, {\bf 77}, 4054 (1996).


\bibitem{auslaender2005spin}
OM~Auslaender, H~Steinberg, A~Yacoby, Y~Tserkovnyak, BI~Halperin, KW~Baldwin,
  LN~Pfeiffer, and KW~West,
 Science, {\bf 308}, 88 (2005).




\bibitem{Kim:2006} 
B. J. Kim, H. Koh, E. Rotenberg, S.-J. Oh, H. Eisaki, N. Motoyama, S. Uchida, T. Tohyama, S. Maekawa, Z.-X. Shen, C. Kim. 
 Nature Physics,  {\bf 2}, 397 (2006). 


\bibitem{Jompol:2009}
Y. Jompol1, C. J. B. Ford, J. P. Griffiths, I. Farrer, G. A. C. Jones, D. Anderson, D. A. Ritchie, T. W. Silk, A. J. Schofield.
 Science, {\bf 325}, 597 ( 2009).


\bibitem{Hulet:2018}
TL~Yang, P~Gri{\v{s}}ins, YT~Chang, ZH~Zhao, CY~Shih, Thierry Giamarchi, and
  RG~Hulet.
 Physical review letters, {\bf 121},  103001 (2018).


\bibitem{Vijayan:2020} Jayadev Vijayan, Pimonpan Sompet, Guillaume Salomon, Joannis Koepsell, Sarah Hirthe, Annabelle Bohrdt, Fabian Grusdt, Immanuel Bloch, Christian Gross.
 Science, 367, {\bf 186} (2020).
 
 \bibitem{Senaratne:2021} R. Senaratne, D. Cavazos-Cavazos, S. Wang, F. He, Y.-T. Chang, A. Kafle,  H. Pu, X.-W. Guan, R. G. Hulet  2021  Science in press.

\bibitem{Lai:1971}C. K. Lai, Phys. Rev. Lett. {\bf 26}, 1472 (1971).
\bibitem{Lai:1973}C. K. Lai, Phys. Rev. A {\bf 8},  2567 (1973).


\bibitem{Affleck1986} I. Affleck, Phys. Rev. Lett. {\bf 56}, 746 (1986).

\bibitem{Cardy1986} J. L. Cardy, Nucl. Phys. B {\bf 270} [FS16], 186 (1986).


\bibitem{Belavin1984} A. A. Belavin, A. M. Polyakov and A. B.
Zamolodchikov, Nucl. Phys. B {\bf 241}, 333 (1984).

\bibitem{Blote1986} H. W. Bl\"{o}te, J. L. Cardy and M. P.
Nightingale, Phys. Rev. Lett. {\bf 55}, 742 (1986).



  \bibitem{Fiete:2007}
  Gregory A. Fiete,
  Review of Modern Physics, {\bf 79}, 801 (2007).
  
  \bibitem{Cheianov:2004} Vadim V. Cheianov and M. B. Zvonarev.
Phys. Rev. Lett.,  {\bf 92}, 176401 (2004).

\bibitem{Takahashi-a}M. Takahashi, Prog. Theor. Phys. {\bf 44}, 899 (1970).

\bibitem{Gu-Yang}C. H. Gu and C. N. Yang, Commun. Math. Phys. 122, 105 (1989).
 
\bibitem{Wadati}T. Iida and M. Wadati, J. Phys. Soc. Jpn, {\bf 77}, 024006 (2008).


\bibitem{Guan:FP}X.-W. Guan, Front. Phys., {\bf 7}, 8 (2012). 

\bibitem{Takahashi:1970-m}M. Takahashi, Prog. Theor. Phys., {\bf 44},  11 (1970).

\bibitem{Schlottmann:1993}P. Schlottmann,  J. Phys. Condens. Matter {\bf 5},  5869 (1993).


\bibitem{Guan:2011-QC} X.-W. Guan and T.-L. Ho, Phys. Rev. A {\bf 84}, 023616 (2011).

\bibitem{Yin:2011}X.  Yin, X.-W.  Guan, M. T. Batchelor and S.  Chen, Phys. Rev. A {\bf 83}, 013602 (2011).

\bibitem{Chen:2014}Y.-Y. Chen, Y.-Z. Jiang,  X.-W. Guan and Q. Zhou, Nat. Comms., {\bf 5}: 5140 (2014). 

\bibitem{He-WB:2016} W.-B. He, Y.-Y. Chen, S.-Z Zhang and X.-W. Guan, Phys. Rev. A, {\bf 94}, 031604(R) (2016).


\bibitem{Larkin1965} A. I. Larkin and Yu. N. Ovchinnikov, Sov. Phys.
JETP {\bf 20}, 762 (1965)

\bibitem{Fulde1964} P. Fulde and R. A. Ferrell, Phys. Rev. {\bf
135}, A550 (1964)

\bibitem{Yang2001} K. Yang, Phys. Rev. B {\bf 63}, 140511(R) (2001)

\bibitem{Liao} Y. Liao \emph{et al.}, Nature \textbf{467}, 567 (2010).

\bibitem{Revelle:2016}M. C. Revelle, J.  A. Fry, B.  A. Olsen, and R. G. Hulet, 
   Phys. Rev. Lett. \textbf{117}, 235301 (2016).

\bibitem{Orso} G. Orso, Phys. Rev. Lett. \textbf{98}, 070402 (2007).

\bibitem{HuiHu} H. Hu, X.-J. Liu and P. D. Drummond, Phys. Rev. Lett. \textbf{98}, 070403 (2007).


\bibitem{Feiguin2007} A. E. Feiguin and F. Heidrich-Meisner, Phys. Rev. B \textbf{76} 220508 (2008).

\bibitem{Rizzi2008}  M. Rizzi, M. Polini, M. A. Cazalilla, M. R.
Bakhtiari, M. P. Tosi and R. Fazio, Phys. Rev. B {\bf 77}, 245105 (2008).

\bibitem{Zhao2008} E. Zhao and W. V. Liu, Phys. Rev. A {\bf 78}, 063605 (2008).

\bibitem{Lee2011} J.-Y. Lee and X.-W. Guan, Nucl. Phys. B {\bf 853},  125 (2011). 

\bibitem{Schlottmann2012} P. Schlottmann and A. A. Zvyagin, Phys. Rev. B {\bf 85}, 205129 (2012).


\bibitem{Peng:2019}L. Peng, Y.-C. Yu and X.-W. Guan, Phys. Rev. B {\bf 100}, 245435 (2019).

\bibitem{Yu:2020}Y.-C. Yu, S.-Z. Zhang and X.-W. Guan, Phys. Rev. Research, {\bf 2}, 043066 (2020).

\bibitem{Yu:2016}Y.-C. Yu, Y.-Y. Chen, H.-Q. Lin, R. A. Roemer, X.-W. Guan,  Phys. Rev. B {\bf 94}, 195129 (2016).



\bibitem{Gruneisen_AdP_1908}
E.~Gr\"uneisen,
Annalen der Physik, {\bf 331}, 211 (1908);  Annalen der Physik, {\bf 344}, 257 (1912).

\bibitem{Batrouni2008} G. G. Batrouni, M. H. Huntley, V. G. Rousseau
and R. T. Scalettar, Phys. Rev. Lett. {\bf 100}, 116405 (2008)

\bibitem{Liu2008} X.-J. Liu, H. Hu and P. D. Drummond, Phys. Rev. A
{\bf 78}, 023601 (2008)

\bibitem{Tezuka2008} M. Tezuka and M. Ueda, Phys. Rev. Lett. {\bf
100}, 110403 (2008)

\bibitem{Frahm1991} H. Frahm and V. E. Korepin, Phys. Rev. B {\bf
43}, 5653 (1991)

\bibitem{Hauke:2016}P. Hauke, M. H. Heyl, L. Tagliacozzo and P. Zoller, Nat. Phys. {\bf 12}, 778 (2016).




\end{thebibliography}
\end{document}